\documentclass[graybox]{svmult}

\usepackage{type1cm}        % activate if the above 3 fonts are
\usepackage{makeidx}         % allows index generation
\usepackage{graphicx}        % standard LaTeX graphics tool
\usepackage{multicol}        % used for the two-column index
\usepackage[bottom]{footmisc}% places footnotes at page bottom

\usepackage{url}
\usepackage{newtxtext}       % 
\usepackage[varvw]{newtxmath}       % selects Times Roman as basic font
\usepackage{booktabs}
\usepackage{fullwidth}
\usepackage{tabularray}

\usepackage{caption}
\usepackage{subcaption}

\usepackage[hide]{commentsnathan}

\newcommand{\Empirical}{\emph{Empirical Methods in Software Engineering}\xspace}

\makeindex             % used for the subject index
\begin{document}

\title*{Teaching Empirical Methods at Eindhoven University of Technology}
\author{Alexander Serebrenik\orcidID{0000-0002-1418-0095} and\\ Nathan Cassee\orcidID{0000-0002-6511-918X}}
\institute{Alexander Serebrenik \at Eindhoven University of Technology, \email{a.serebrenik@tue.nl}
\and Nathan Cassee \at Eindhoven University of Technology, \email{n.w.cassee@tue.nl}}
\maketitle

\abstract*{In this chapter, we share an experience report of teaching a master course on empirical research methods at Eindhoven University of Technology in the Netherlands. 
The course is taught for ten weeks to a mix of students from different study programs and combines both practical assignments with a closed-book exam.
We discuss the challenges of teaching a course on research methods and explain how we address these challenges in the course design. Additionally, we share our lessons learned and the do's and don'ts we learned over several iterations of teaching the course. 
}

\abstract{
%\todo{Check me! roughly 100 words, max is 200.} 
In this chapter, we share an experience report of teaching a master course on empirical research methods at Eindhoven University of Technology in the Netherlands. 
The course is taught for ten weeks to a mix of students from different study programs and combines both practical assignments with a closed-book exam.
We discuss the challenges of teaching a course on research methods and explain how we address these challenges in the course design. Additionally, we share our lessons learned and the challenges we encountered over several iterations of teaching the course. }

\section{Introduction}
Empirical methods in software engineering are taught in numerous institutions world wide, from Canada\footnote{e.g., at University of Victoria \url{https://emse-uvic.github.io/index.html} or at University of Toronto \url{https://www.eecg.utoronto.ca/~shuruiz/teaching/ECE1785-2021W/}.} and the United States\footnote{e.g., at Carnegie Mellon University \url{https://bvasiles.github.io/empirical-methods/} or at North Dakota State University \url{https://catalog.ndsu.edu/search/?P=CSCI\%20848}} 
to Germany\footnote{e.g., at Freie Universit\"{a}t Berlin \url{https://www.mi.fu-berlin.de/w/SE/VorlesungEmpirie2022} or Universit\"{a}t des Saarlandes \url{https://cms.sic.saarland/empse_ws_2223/}},
Sweden,\footnote{e.g., at Gothenburg University \url{https://www.gu.se/en/study-gothenburg/empirical-software-engineering-dit246}}
Norway\footnote{e.g., at NTNU \url{https://www.ntnu.edu/studies/courses/IT3010#tab=omEmnet}}, and from Brazil\footnote{e.g., at Universidade Tecnol\'{o}gica Federal do Paran\'{a}  \url{https://www.utfpr.edu.br/cursos/coordenacoes/stricto-sensu/ppgcc-cm/area-academica/disciplinas/engenharia-de-software-empirica} or at Universidade Federal do Acre \url{https://portal.ufac.br/ementario/disciplina.action;jsessionid=ED0C1B1B9875484A460ECF2EC64D358D?d=22091}} and Uruguay\footnote{e.g., at Universidad ORT \url{https://fi.ort.edu.uy/92839/33/}} to Australia\footnote{e.g., at the University of Adelaide\url{https://www.adelaide.edu.au/course-outlines/108290/1/sem-1/}}. 
While bearing similar names, these courses are different: they target different student populations, engage with larger or smaller groups, and cover different selections of topics. 

This experience report reflects on three years of teaching empirical software engineering in the master's program at Eindhoven University of Technology (TU/e), The Netherlands. 
To accompany this chapter we also share all teaching materials we have designed for the course.
We hope that this report helps colleagues who are teaching similar courses elsewhere.

\section{Background: Master Courses at Eindhoven University of Technology}
\subsection{Student Demographics}
Eindhoven University of Technology offers 24 different master programs including Computer Science and Engineering, Data Science and Artificial Intelligence, and Human-Technology Interaction.
In addition to a limited number of mandatory courses determined by each program, students are free to take any course of their liking.
Most of the courses offered, including \Empirical, are indeed open for students of any master's program; the aforementioned programs, however, explicitly list \Empirical as one of the recommended courses or even prescribe it to subgroups of their students.
Most students registered for the 2023/2024 edition of \Empirical follow either the Computer Science and Engineering master program (40) or Data Science and Artificial Intelligence (21). 
The remaining students come from such master programs as Human-Technology Interaction (2 students), Embedded Systems, Industrial and Applied Mathematics, Artificial Intelligence Engineering Systems, and pre-master program Computer Science and Engineering (1 student each).
Furthermore, the number of students increases each year: from 41 in 2020 to 51 in 2021, 60 in 2022, and 67 in 2023.
Hence, when designing the course, we had to take into account \emph{a relatively large number of students} taking the course.

Moreover, diversity of master programs implies  \emph{diversity of the students' educational backgrounds}: e.g., while data science students had limited exposure to basic software engineering practices, they have ample experience with training machine learning classifiers and quantitative data analysis.
The situation is reversed for computer science students. 
Students in the Human-Technology Interaction master program often have a bachelor's degree in psychology and, hence, are experienced with conducting quantitative surveys and designing interventions but not with analysing data from online platforms. 
Finally, diversity in educational background does not only refer to the academic discipline but also to the previous education: while some students hold a bachelor degree from more scientifically-oriented institutions such as Eindhoven University of Technology itself, others hold a bachelor degree from a Dutch university of applied sciences, which is much more focused on the engineering practice.
Such students are often much better software developers than their peers trained at scientifically oriented institutions but are less familiar with more abstract, more conceptual ways of thinking.

\subsection{Schedule}
All master programs are designed for two academic years: students are expected to start their education in September and graduate in August of the following \textcolor{black}{academic} year. 
In practice, however, students sometimes take longer than two years to complete their degrees for various reasons.

Every course is offered during one teaching period, consisting of seven regular teaching weeks, one week that can be used to compensate for the teacher's absence and two exam weeks. 
During a regular week each course is allocated two sessions, one in the morning and another one in the afternoon. 
Each session consists of two academic hours (45 minutes) separated by a 15-minutes break. 
Every course is considered to be 5 ECTS worth.
ECTS, European Credit Transfer and Accumulation System, express the ``volume of learning based on the defined learning outcomes and their associated workload''~\cite{ECTS}.
A full-time study during one academic year corresponds to 60 ECTS. 
According to the Dutch norm, 1 ECTS corresponds to 28 hours of study, i.e., an average student is expected to spend 140 hours studying for a single course.
Since the number of contact hours is limited to 28 (seven weeks, two meetings per week, two hours per meeting), \emph{the lion's share of the course work should be organized as a self-study}. 

Since \Empirical is an elective subject for most students, they can take it either during their first year or during their second year.
However, in practice \Empirical is usually taken at the beginning of the masters program, when students are not yet working on any research projects. 
At the same time, it is supposed to prepare the students for conducting a research project leading towards their master thesis. 
This imposes the challenge of teaching the students how to conduct empirical research without giving them the feeling that they are practicing a skill that they do not need. 

\subsection{Assessment}
Teachers of courses at Eindhoven University of Technology are free to choose between a written exam, an oral exam, an assessment based on homework assignments, or any combination of those.
In case the assessment is based on group projects, part of the grade should be determined based on individual contributions.
A written exam can take up to three hours and the resit is organized during the evening hours after the subsequent teaching period, i.e., ca. ten weeks after the first attempt.
Under special circumstances upon student's request a faculty-level committee can ask the teacher to organise another resit; however such resits are uncommon.
Most teachers opt for a combination of an assessment based on homework assignments, often organized as group assignments, and a written exam.
Also, here, diversity of educational backgrounds plays a role.
\textcolor{black}{Students holding a bachelor's degree from the Netherlands can be expected to be familiar with group assignments: e.g., most of the courses in the computer science bachelor of Eindhoven University of Technology have some form of group projects.
However, students holding a bachelor's degree from abroad have not necessarily had the same exposure to group projects.}

\textcolor{black}{At the end of the course students receive a course grade that is an integer between 1 and 10 with 6 being the minimal passing grade. 
Usually, the final course grade is determined as the weighted average of the grades for individual course components such as the exam and the homework assignments, rounded to the nearest integer.
Very high grades course grades exceeding 9 are uncommon.
Grades for the individual course components are between 1 and 10 and can have one digit after the decimal point, e.g., 7.4. 
}

\subsection{Resources}
The teaching staff of a master course usually consists of one or two teachers, rarely supported by student assistants.
For \Empirical, the only teachers involved are the authors of this chapter.

Finally, students and teachers are encouraged to use Canvas LMS, a learning management system, and ANS-Delft, an assessment platform designed to support paper, digital, and hybrid exams.

\subsection{Summary}
In summary, the main educational challenge addressed in this course design is:  
\textbf{How to teach, in ten weeks, a self-study-driven, realistic course on empirical software engineering research methods to a group of 50-70 students with no previous research experience and diverse backgrounds while considering the limited resources available.} 

\section{\Empirical}

In this section, we discuss the learning objectives of the course, the teaching activities, and the assessment. 

\subsection{Learning objectives}
As the first step towards designing the course, we have formulated the following learning objectives: 

\begin{enumerate}
    \item[\textbf{LO1}] Students will be able to formulate and motivate research questions pertaining to software engineering, and identify research questions that can and cannot be answered by means of empirical research.
    \item[\textbf{LO2}] Students will be able to execute a valid and trustworthy empirical study in Software Engineering given 
        a state-of-the-art dataset. 
    \item[\textbf{LO3}] Students will be able to evaluate empirical studies in Software Engineering using instruments accepted in the field, and be able to 
    to identify threats to validity and problems with trustworthiness. 
    \item[\textbf{LO4}] Students will be able to describe the results of empirical studies to professionals not familiar with 
     academic research.
     \item[\textbf{LO5}] Students will be able to comprehend the research methods used for empirical studies in Software Engineering.
\end{enumerate}

We strongly believe that these learning objectives capture the most important aspects of conducting empirical research in software engineering.
\textbf{LO1}, \textbf{LO2}, and \textbf{LO3} address the steps inherent in any empirical research and, subsequently, the skills the students will need when conducting empirical research either as part of their study or as part of their further career. 
However, before students can conduct empirical studies, they must understand the research methods commonly used in empirical studies (\textbf{LO5}).
Finally, \textbf{LO4} relates to the societal obligation of researchers to convey the research findings to practitioners.

\subsection{Course organisation}
To achieve the learning objectives above, with the aforementioned educational challenges in mind, we have offered an overview course showcasing several different empirical methods and their applications to software engineering research.
The teaching activities of \Empirical consist of a mix of lectures, workshops, and coffee-hours.
As customary, \emph{lectures} are used to teach students the theoretical foundations of empirical research methods (\textbf{LO5}). 
In contrast, the \emph{workshops} are more interactive and are intended for students to work on bridging the gap between a more abstract presentation of the research methods and their application in the practice of software engineering research (\textbf{LO1}--\textbf{LO4}).
The \emph{coffee-hours} are scheduled during the regular lecture hours and are a platform for students to ask questions about the course and the assignments.
Our decision to allocate time to workshops and coffee hours implies that we had to limit the number of topics that could be covered during the course.
When selecting the topics for the lectures, we decided to include topics that are orthogonal to the specific choice of a research method (e.g., stating research questions and choosing sampling techniques), as well as topics showcasing the diversity of empirical software engineering research methods (e.g., interviews vs. repository mining, quantitative analysis vs. qualitative analysis).
The overview of the course schedule for the 2022 %and 2023 
edition can be found in Table~\ref{tab:lec-schedule-2022}.

\begin{table*}
    
    \caption{Course schedule for the 2022 edition of \Empirical.} 

\begin{tabular}{llp{7.5cm}}\toprule

    \textbf{When} & \textbf{What} & \textbf{Topic} \\
    \midrule
    Nov 15\textsc{th} 13:30 & Lecture & Course introduction. Empiricism and Rationalism  \\
    Nov 17\textsc{th}  08:45 & Lecture & Research Questions and Research Strategies  \\ \midrule

    Nov 22\textsc{nd} 13:30 & Guest Lecture & Empirical Methods in the Industry \\
    Nov 24\textsc{th} 08:45 & Lecture & Sampling  \\
    Nov 25\textsc{th} 23:59 & Deadline & 1\textsc{st} deadline for Design a study \\ \midrule

    Nov 29\textsc{th} 13:30 & Workshop & How to read an empirical paper \\
    Dec 1\textsc{st} 08:45 & Lecture & Interviews \& Surveys \\ \midrule

    Dec 6\textsc{th} 13:30 & Lecture & Mining Software Repositories I \\
    Dec 8\textsc{th} 08:45 & Workshop & Mining Software Repositories \\ 
    Dec 9\textsc{th} 23:59 & Deadline & 2\textsc{nd} deadline for Design a study \\ \midrule

    Dec 13\textsc{th} 13:30 & Lecture & Quantitative Analysis  \\
    Dec 15\textsc{th} 08:45 & Lecture & Qualitative Analysis \\ 
    Dec 16\textsc{th} 23:59 & Deadline & 1\textsc{st} deadline for Describe a study \\ \midrule

    Dec 20\textsc{th} 13:30 & Coffee-hour & Feedback and assignment Q\&A session \\
    Dec 22\textsc{th} 08:45 & Lecture & Advanced Repository Mining \\
    Dec 23\textsc{rd} 23:59 & Deadline & 3\textsc{rd} deadline for Design a study \\ \midrule

    Jan 10\textsc{th} 13:30 & Coffee-hour & Feedback and assignment Q\&A session \\
    Jan 12\textsc{th} 08:45 & Lecture & Threats to Validity. Trustworthiness.  \\
    Jan 13\textsc{th} 23:59 & Deadline & 2\textsc{nd} deadline for Describe a Study \\ \midrule
    
    Jan 17\textsc{th} 13:30 & Workshop  & Recap \& Threats to Validity. \\
    Jan 19\textsc{th} 08:45 & No lecture &  \\ \midrule

    Jan 31\textsc{st} 09:00 & Exam & Final examination \\
    Feb 3\textsc{rd} 23:59 & Deadline & Final deadline for Design a study \\ \midrule

    Apr 12\textsc{th} 13:30 & Exam & Resit \\ 
    \bottomrule
\end{tabular}

\label{tab:lec-schedule-2022}
\end{table*}

\subsection{Lectures}

During the first lecture of \Empirical we introduce the course structure and discuss the assignments and deadlines. 
This helps clear up any confusion students might have about the course and the assignments and ensures that students know what is expected of them.
Next, we introduce the fundamental differences between empiricism and rationalism (based on the article of Ralph~\cite{Ralph2018Two}) and discuss several examples of empirical software engineering papers~\cite{Khalid2015What,Krueger2020Neurological,Meyer2021Today,Qiu2019Going,Zhou2022CrossCompany}. 
We have chosen these papers because they show a broad spectrum of software engineering problems being addressed and a wide variety of data collection and analysis methods employed.
This introductory lecture is followed by a lecture on research questions and research strategies, based on the classical chapter by Easterbrook et al.~\cite{Easterbrook2008Selecting} and the article of Stol and Fitzgerald~\cite{Stol2018ABC}.

The second week starts with a guest lecture, in which guests from industry discuss how they apply empirical research methods. 
To prevent guest speakers from going out of scope we generally tend to invite contacts that we already know, additionally, we clearly communicate the goal of the guest-lecture to the speakers. 
The strategy of Eindhoven University of Technology for 2030 mentions a need for the practical application of theoretical knowledge.  
This is why we use the guest lectures to explain to students that applying empirical methods is not limited to ``the ivory tower of academia''.
In 2022, we allocated two hours for a guest lecture, but in 2023, we decided to shorten it to one hour to make space for a lecture on Design Science.
Indeed, in 2022, we have observed that while the lion's share of the course was dedicated to providing students with a means of understanding software engineering phenomena, in their follow-up master projects, they are often tasked with designing tools and evaluating their performance.
To address the needs related to designing tools, and to help students reason about the novelty and relevance of their work, we included a one-hour lecture on design science based on the article by Engstr{\"{o}}m et al.~\cite{Engstrom2020How}. 
In the second week we also teach a lecture on the topic of sampling following Baltes and Ralph~\cite{Baltes2022Sampling}.

Lectures of the third and fourth weeks are dedicated to data collection techniques: interviews and surveys, based on such papers as the work of Seaman~\cite{Seaman2008Qualitative}, Hove and Anda~\cite{Hove2005Experiences}, and Strandberg~\cite{Strandberg2019Ethical}; and mining software repositories including discussion of the ``promises and perils'' associated with repository mining~\cite{Bird2009Promises,Kalliamvakou2016Indepth}. 
Lectures of the fifth and the sixth weeks focus on data analysis: 
quantitative, bring students up-to-speed with such statistical methods as regression discontinuity design~\cite{Imbens2008Regression} and mixed-effects modeling~\cite{Bates2015Fitting};  qualitative, making the students aware that (post-)positivism is not the only valid paradigm in software engineering research based on the discussion by Melegati and Wang~\cite{Melegati2021Surfacing}, and introducing the basic notions related to grounded theory following Hoda~\cite{Hoda2022STGT}; as well as on specific issues related to analysis of data from software repositories.

The last lectures cover the topics related to threats to validity following Wohlin et al.~\cite{Wohlin2012Experimentation} and trustworthiness in qualitative studies following Lincoln and Guba~\cite{Lincoln1985Naturalistic}. 
Finally, the course is concluded with a recapitulation lesson.

\subsection{Workshops}
The first edition of the course (2020) consisted solely of lectures.
Due to the COVID-19 pandemic, the lectures have been video recorded, which is why we experimented with a flipped classroom design in 2021. 
In this design, we requested that students watch the recorded lectures at home, and we organized workshops for every teaching session.
This was not appreciated by the students, as they missed a more traditional way of transferring knowledge. 
The recorded lectures were ill-suited as a sole way of supporting self-study, and organizing two workshops per week has incurred an enormous amount of work for us.
Additionally, we found that not all topics were equally suitable for 90-minute workshops: e.g., in the workshop on interviews we found that 90 minutes is simply too brief to let students practice with the drafting and practicing of semi-structured interviews.
This is why, from 2022 onward, we have decided to revert to the more traditional lecture-based setting, allocating three teaching sessions to workshops. 
As shown in Table~\ref{tab:lec-schedule-2022} the workshops are dedicated to reading empirical papers, mining software repositories, and threats to validity. 

\subsubsection{Workshop: Reading Empirical Papers}
During the reading workshop, the students are introduced to academic publishing (e.g., publication venues and structure of an empirical paper). 
Then, they are given 15 minutes to scan an empirical paper assigned to them and prepare a two-minute-long pitch in which they answer the following questions:
\begin{itemize}
\item What problems do the authors try to
address? 
\item Who benefits from the research of the
authors? 
\item What strategy following Stol and Fitzgerald~\cite{Stol2018ABC} do they use?
\item What research methods in the strategy do they
use?
\item Is this paper a Design Science paper? If so, what is the design science
component? (2023)
\item What might have invalidated the results of
the authors?
\item What did the authors find?
\end{itemize}
For this task we used papers by Moldon et al.~\cite{Moldon2021How}, Vidoni~\cite{Vidoni2021Evaluating}, Wan et al.~\cite{Wan2021ML} and Danilova et al.~\cite{Danilova2021Do}.
The students then pitch the papers to each other and compare their insights.
We try to facilitate a class discussion after each pitch, so we ask the students who also read the paper to respond to the pitch.   

In the second part of the workshop, we introduce students to the idea that a single published paper might report the results of two or more individual empirical studies. 
Therefore, we ask the students to read a different paper (either by Begel and Zimmermann~\cite{Begel2014Analyze} or by Ralph et al.~\cite{Ralph2020Pandemic}) in more detail and indicate how many individual studies the paper describes. 
For each study, we then ask them to identify the research question, the methodology, the sampling strategy used and the conclusions of the studies. 
The students are asked to discuss their findings with their neighbors.

\begin{longtblr}
    [
        label={table:lesson-plan},
        caption={Lesson plan for the workshop on Mining Software Repositories.\\}
    ] {
        width=0.98\textwidth,
        colspec={X[0.7,l] X[1,l] X[1,l] X[6.5,l]},
        hline{1,Z} = {1.5pt},
        hline{2} = {1.2pt},
        hline{3-Y} = {0pt},
    }
    %\toprule
    \textbf{Time (min)} & \textbf{Topic} & \textbf{Resource} & \textbf{Content}  \\*
     5 & Recap & Slides & Welcome students, open the lecture. Start by recapitulating the theory on research strategies and 
     the lecture on MSR. Explain how the material of this workshop allows students to apply this material  \\
     10 & Gap \& Tool & Slides & Briefly summarize two studies that use Stack Overflow as dataset,  summarize the research gap introduced in these two papers, and introduce the publicly available StackOverflow dataset students will use for the remainder of the workshop. \\
     5 & Think & -- & Ask students to draft research questions based on the research gap individually. Remind them of the material on research strategies, and how to draft research questions. \\
     5 & Share & Excel & Ask students to share their research questions with their neighbors and evaluate their partner's research question concerning motivation and relevance. Finally, ask some pairs to share their research questions with the group, and give feedback on them so that students can see how I evaluate research questions. Finally, ask students to share their research questions in the shared Excel sheet. \\
     5 & Tips & Slides & Give students a set of tips and best practices for conducting MSR research. Tell them that for the rest of the workshop they will work on answering their drafted research questions using the Stack Overflow dataset. \\
     15 & Work & -- & Students work on their research questions, a teacher walks around to ask pairs of students how they are progressing and if they have any questions. \\
     15 & Break & -- & -- \\
     20 & Work & -- & Students work on their research questions. Where needed a teacher supports students \\
     5 & Update & -- & Ask students who have already obtained results to share their results with the class, and discuss their approaches and their queries. Ask students to reflect on their approaches and whether and how they have followed the guidelines.  \\
     15 & Work & -- &  Students work on their research questions. We try to support those students that are having more trouble with the material, so that at the end of the workshop hopefully everyone has some results. \\
     5 & Wrap-up & Slides & Wrap-up the workshop, summarize and explain how students can use the material of this lecture for their assignment. Announce the topic of the next lecture. \\
      %\bottomrule
\end{longtblr}

\subsubsection{Workshop: Mining Software Repositories}

The workshop occurs after a lecture on Mining Software Repositories (MSR).
The workshop aims to teach students how to apply the material from the lecture on MSR (Promises \& Perils) to a specific instance of a dataset. 
Additionally, we use this workshop to recapitulate the material on drafting and evaluating research questions, as this is something that the students need for their assignment \emph{Design a Study} and are known to have difficulties with. 
The learning goals of the workshop are the following: 
\begin{itemize}
    \item Students are able to formulate research questions based on a dataset and a research gap.
    \item Students are able to evaluate research questions with respect to feasibility and motivation. 
    \item Students are able to apply the material of MSR to collect data from a software engineering dataset.
\end{itemize}

The workshop opens with a brief slide deck to recapitulate relevant theory for the workshop and to introduce the dataset (Stack Exchange data explorer\footnote{\url{https://data.stackexchange.com/}}). 
Note that while the lecture on MSR covered the platforms Git, and GitHub, the workshop uses Stack Exchange. 
We purposefully chose Stack Exchange, such that students have to reason about a new platform. 
In the presentation. we also cover a research gap that students will work-on during the workshop.
Additionally, students are generally familiar with Stack Overflow, and we expect them to have some familiarity with SQL. 

Table \ref{table:lesson-plan} lists the lesson plan of the workshop, in the workshop we purposefully apply Rosenshine's principles of instruction~\cite{Rosenshine:1986}.
When opening the workshop, we ground the material of the workshop by recapitulating the material of the lecture on MSR and we explain how the material of this workshop relates to the research strategies of Stol and Fitzgerald~\cite{Stol2018ABC} discussed in Lecture 2. 
Furthermore, we alternate the periods where a teacher is talking with periods where the students are engaging with the material themselves.

\subsubsection{Workshop: Threats to Validity}

During the threats to validity workshop, we let students draft and discuss threats to validity according to the model of Wohlin~\cite{Wohlin2012Experimentation}. 
We start the workshop by briefly recapitulating the theory on threats to validity. 
After recapitulating we give a students a small interactive exercise: We show them a snippet in which a threat to validity is discussed, and we ask them whether the threat to validity was a threat to \emph{Internal}, \emph{External}, \emph{Construct} or \emph{Conclusion} validity. 
Afterward, we host a small discussion in which we ask students to explain their answers and discuss their reasoning. 
We use this moment to re-iterate how different threats to validity are distinguished from each other and their definitions. 

Continuing on, we briefly explain some common pitfalls in threats to validity.
For instance, we explain to students that threats to validity are not an excuse to do nothing: 
While working on their projects, they cannot make faulty choices and describe them as threats to validity. 

During the final 45 minutes, we end with one large, practical exercise. We give students the paper of Kinsman et al. ~\cite{Kinsman:2021} in which we redacted the threats to validity section. 
We ask students to read the paper, identify the studies that are part of the paper, and then identify any threats to validity present in the paper. 
In a way, this exercise is the culmination of most of the knowledge we expect students to master: 
They need to read a paper, identify the methodologies used within it, and discuss any relevant shortcomings of the authors that might invalidate or threaten conclusions. 
Furthermore, the paper of Kinsman et al. combines both qualitative and quantitative methods, giving us a chance to discuss a wide range of threats to validity. 
Usually, while students are reading the paper, we walk around and engage in conversation with them. 
These one-on-one conversations allow us to guide students in their thinking and discuss their reasoning in a bit more detail. 
Finally, we wrap up the workshop with a brief central discussion; we invite students to share their threats and encourage students to give feedback on each other's threats. 
%\todo{Check this section}

\subsection{Assessment}
\label{sec:assessment}
Similarly to other courses at our university we opt for a combination of a written exam and homework assignments. 
The final grade for the course is the weighted average of the \emph{assignment grade} (70\%) and 
the exam (30\%). 
Additionally, to pass the course, an individual student's exam and assignment grades should both be 5.0 or higher.
This requirement ensures that students who fail either the exam or both assignments cannot pass the course.
The \emph{assignment grade} is the weighted average of the Design a Study assignment (70\%) and the Describe a Study assignment (30\%).

\begin{table}[tbh]
    \begin{tabular*}{\linewidth}{@{\extracolsep{\fill}} lrrrrr r}
    \toprule
    \textbf{Assessment} & \textbf{LO1} & \textbf{LO2} & \textbf{LO3} & \textbf{LO4} & \textbf{LO5} & \textbf{Weight} \\
    \midrule
      Design a Study  & 10\% & 28\% & 7\% & & 5\% & 50\% \\
      Describe a Study   & &  &  & 20\% & & 20\%  \\ 
     Exam  & & 12\% & 12\% & & 8\% & 30\% \\
      \midrule 
      Total & 10\% & 40\% & 19\% & 20\% & 13\% & 100\% \\ 
      \bottomrule
    \end{tabular*}
    \caption{Weighting of each assessment moment and the learning objectives each assessment covers.}
  %  \setfloatalignment{t}
    \label{table:assessment-matrix}
  \end{table}

\subsubsection{Design a Study}

Design a Study is a group assignment designed for groups of four students.
During Design a Study, students are asked to take an existing dataset and use it to conduct an empirical study.
We have paid special attention to selecting papers with dataset descriptions. 
We let students choose between different kinds of data, e.g., data about bugs and data about developer communication; because we want to reduce the risk of plagiarism and give students the freedom to choose a dataset they think is interesting.
The writing should be easily accessible for students with limited software engineering knowledge, the data should be stored either as comma-separated values or as a relational database, and the dataset should not be too large. 
The screening of datasets allows our students to focus more on applying the course material, as in 2021, unfortunately, working with a large No-SQL database was too challenging for many students.
Students spent much time and effort trying to master the No-SQL query language and not on empirical research. 

In 2022, we have suggested the students the following MSR 2022 datasets:
\begin{itemize}
    \item A Large-scale Dataset of (Open Source) License Text Variants~\cite{Zacchiroli2022Dataset} \footnote{\url{https://hal.archives-ouvertes.fr/hal-03624198/document}}
    \item An Alternative Issue Tracking Dataset of Public Jira Repositories~\cite{Montgomery2022Dataset}\footnote{\url{https://arxiv.org/pdf/2201.08368.pdf}}
    \item A Time Series-Based Dataset of Open-Source Software Evolution~\cite{Sousa2022Dataset}\footnote{\url{https://drive.google.com/file/d/1svIexRcWrXBhb_pCVhfs7yKRzGsyaBpV/view}}
    \item A Versatile Dataset of Agile Open Source Software Projects~\cite{Tawosi2022Dataset} \footnote{\url{https://solar.cs.ucl.ac.uk/pdf/tawosi2022msr.pdf}}
    \item DaSEA – A Dataset for Software Ecosystem Analysis~\cite{Buchkova2022Dataset} \footnote{\url{https://itu.dk/~ropf/blog/assets/msr2022.pdf}}
    \item DISCO: A Dataset of Discord Chat Conversations for Software Engineering Research ~\cite{Subash2022Dataset}\footnote{\url{http://olgabaysal.com/pdf/MuthuSubash_MSR2022_DataShowcase.pdf}}
    \item The OCEAN mailing list data set: Network analysis spanning mailing lists and code repositories~\cite{Warrick2022Dataset} \footnote{\url{https://arxiv.org/pdf/2204.00603.pdf}}
\end{itemize}

Each year we replace some of the datasets, based on the popularity of the datasets among the students. For the academic year 2023 we replaced the first two datasets with: 

\begin{itemize}
   \item A Dataset of Bot and Human Activities in GitHub~\cite{Chidambaram2023Dataset} \footnote{\url{https://decan.lexpage.net/files/MSR-2023.pdf}}
    \item GitHub OSS Governance File Dataset~\cite{Yan2023Dataset} \footnote{\url{https://zenodo.org/records/7530768}}
    \item GIRT-DATA: Sampling GitHub Issue Report Templates~\cite{Nikeghbal2023Dataset} \footnote{\url{https://arxiv.org/pdf/2303.09236.pdf}}
\end{itemize}

In this assignment, students are tasked with designing novel research questions, motivating and contextualizing their research questions, positioning their work with respect to the existing literature, designing and describing an appropriate methodology, reporting the results, and reasoning about the validity and trustworthiness of their work. 
With this assignment, we aim to cover the learning objectives of \Empirical that are related to the design, execution, and evaluation of an empirical study (\textbf{LO1}, \textbf{LO2}, \textbf{LO3}).

In the first editions of the course, we observed that students had difficulties coming up with good research questions or picking the right methodology.
This is why we have decided to split the assignment into four parts with four different deadlines: students are encouraged but not required to submit partial versions of the report by the first three deadlines.
We provide them with feedback on the novelty of the research questions, the feasibility of the proposed study, the writing, the relevance of the research questions, the motivation of the work, and the methodology. 
Where possible we try to give our feedback in the form of open, reflective, questions. 
We do not grade the submissions: this allows students to make mistakes, learn from these mistakes without being penalized, and iteratively improve their work. 
Since the same assessment criteria are used to grade the final submission, the students get an early indication of their learning progress.

Specifically, by the first deadline, the students should select three of the above dataset papers, summarize each paper in a single paragraph, and propose four research questions that one can answer based on one (or more) of the datasets. 
By the second deadline, the student should submit a draft of their final report, including an introduction, related work, and methodology. 
The study design should address two research questions based on one or two of the datasets. 
By the third deadline, the students should revise the report based on the feedback received so far and sketch the study results.
The fourth and final deadline of the assignment is the full submission of the report, including results, discussion, threats to validity, and conclusions.
Since \emph{Design a Study} is a complex assignment, the final deadline is at the very end of the course, after the exam (see Table~\ref{tab:lec-schedule-2022}).
In this way we create a gap of four weeks between the last formative and final deadlines. 
This ensures that the students have four weeks to work on the assignment, implement our feedback, and finalize their report. 

\subsubsection{Describe a Study}
\label{sec:describe}
Describe a Study is an individual assignment.
The goal of Describe a Study is to teach students how to accurately communicate the findings from the academic literature in a way that is accessible to a broad audience. 
Graduates will likely be working in teams where not everyone is familiar with reading and disseminating academic literature, and they should be able to build bridges between scientific research and industrial practice.
The assignment covers \textbf{LO4}; it was inspired by the ``It Will Never Work in Theory'' project of Greg Wilson\footnote{\url{https://neverworkintheory.org/category/}} and the Practitioners' Digest initiated by IEEE Software in 2015~\cite{Spinellis2015Introductions}. 

Since students are usually not familiar with writing a summary of a paper for a broader audience, in 2022, we have chosen to have two iterations of the assignment. 
For each iteration, we have selected several recently published empirical software engineering studies that do not explicitly list implications for practitioners.
We restricted the number of students summarizing the same paper to seven; we have offered choice of 11 papers for the first iteration~\cite{AlAmin2021Empirical,Alfadel2021Use,Ciniselli2021Empirical,Eng2021Revisiting,Galappaththi2022Does,Grotov2022LargeScale,Hora2021Googling,Kinsman2021How,Kochanthara2022Painting,Vidoni2021SATD,Young2021Which}, and 7 papers for the second iteration~\cite{Fang2022Slick,Gallaba2022Lessons,Hu2022Practitioners,Liang2022Understanding,Miller2022Did,Santos2022Grounded,Xiao2022Recommending}.
Like Design a Study, the first iteration is not mandatory or graded, instead we provide feedback to help you improve the students' writing. 
For the second iteration, the students were requested to submit a summary of a different paper.

The feedback we have received from the students has indicated, however, that their chances to improve the summary were limited due to switching to a different paper in the second iteration.
This is why, in 2023, while keeping the two-iterations principle, we adjusted the assignment so that students could work on the same scientific paper for both iterations.
Specifically, for \underline{the first iteration}, the students select a paper and write a summary or script for the two-minute-long video pitch. 
The students are expected to remember that practitioners are not familiar with academic research or the methodologies used by software engineering researchers.
Furthermore, the students should be aware that while researchers are interested in the exact methodology, practitioners can be expected to care less about how a study was conducted but instead about what the study means for their work.
For \underline{the second iteration}, the students record the pitch as if they have been invited to speak at a large developer conference.
To record the video the students are free to use any tool they like, however, in the video the students' face should be visible. 
%We require this both to prepare students to giving public talks as well as to facilitate understanding
Additionally, the students are free to use any supporting material they like, such as slides or images, however, these should not distract too much from the pitch itself.

The grading criteria for the pitch are the quality of the content and delivery as well as its relevance for practitioners:
\begin{itemize}
\item \emph{Content}: Does the pitch contain a clear and accurate summary of the research paper? 
Does the pitch contain a clear description of the findings of the paper?
\item \emph{Delivery}: Is the pitch delivered in a clear and concise manner? 
Is the pitch delivered in a way that is engaging for the audience? 
Does the pitch stay within the time limit?
\item \emph{Relevance}: Does your pitch highlight the relevance of the research
for practitioners? 
Does it accurately address the question: Why should practitioners care about this research?
\end{itemize}
%\todo{Details}

\subsubsection{Exam}
\label{sec:exam}
The 2021 exam consisted mostly of multiple-choice questions to ensure that grading of the exam could be done efficiently.
However, after speaking to students at the end of the exam and inspecting the grades, we found out that even good students sometimes understood the material but interpreted questions differently than intended.
Moreover, by definition, such an exam provides very limited insights into the way students reason about a study design

Hence, since the 2022 edition of \Empirical we opt for a 90 minutes closed book exam. 
The exam covers learning objectives \textbf{LO3} and \textbf{LO5} and consists of a mix of open and multiple-choice questions. 
Five multiple-choice questions (4 points each) are intended to check the understanding of the basic notions of empirical software engineering research (learning outcome \textbf{LO5}): e.g., one of the questions asked the students to identify which of the following aims would justify conducting a laboratory experiment: building a new theory based on the existing ones,  establishing causal relations, gaining profound understanding of the study participants' experiences or opinions, or understanding the phenomenon under study in the most realistic context. 
Next, we sketch the context of a scientific study and a corresponding research question and ask the students to design a study methodology that would allow a researcher to answer the research question. 
To help the students, we ask a series of subquestions related to the study design, i.e., motivate the choice for the research strategy among those described by Stol and Fitzgerald~\cite{Stol2018ABC}, describe a data collection method, describe and motivate the sampling approach, describe and motivate the data analysis approach. 
In total, this group of questions amounts to 40 points.
For the final part of the exam we provide the students with a four-page empirical software engineering paper with the Threats to validity section censored. 
The students should read the paper and propose two viable and distinct threats to validity. 
For each threat to validity, they should explain which conclusions of the study would be invalidated by the threat and how, and classify the threat as a threat to construct, internal, external or conclusion validity. 
The threats described should belong to two different types of threats to validity, i.e., if the first threat to validity belongs to the category ``external'', the second threat you describe cannot belong to the category ``external''. 
This part of the exam is also scored on a scale of 40 (9 points for the description of each one of the threats, 9 points for the explanation of what conclusions are invalidated and how, and 2 points for the classification of the threats in the corresponding category).

\subsection{Limitations and Implications}

The choices made in the course design described in this section were partly made based on the following limitations:

\begin{itemize}

    \item \textcolor{black}{\emph{Course duration}: Because of the ten-week-long teaching period at Eindhoven, there is not enough time to have students write a \emph{publishable} scientific study during the course. 
    While we would consider this to be the most appropriate practical assignment for the course, it is unfortunately infeasible given the very short timeline of the course: The Design a Study reports written by students often focus on less-relevant research questions, and because of the ten-week teaching period there is no time to give students enough time to rework their own research questions. 
This being said we have co-authored an MSR mining challenge paper~\cite{Moharil2022Between} with a group of students from the 2021 edition of \Empirical based on their Design a Study report.
Unfortunately, this is no longer an option for more recent course editions due to the MSR mining challenge deadlines being moved from February to December and the course being taught from mid-November to early February.}

\item  \textcolor{black}{\emph{Technological Knowledge}: 
    Students participating in \Empirical are diverse with respect to their study program and prior knowledge. Because of this, some students are experienced programmers and data analysts. However, there are also students who have very little programming experience, and this limits the topics we can cover during more technical workshops, such as the one on MSR. 
    Additionally, in one of the previous editions, we used a dataset stored in MongoDB: querying a non-relational database turned out to be too difficult for some of the students hindering their ability to ask queries interesting from the software engineering perspective.
    Therefore, we have discontinued using this dataset, and instead, we let students pick from a more diverse set of datasets for the \emph{Design a Study} assignment. }

\item \textcolor{black}{\emph{Available Resources}: 
    It is not ideal to use a closed-book exam to test the course's high-level learning objectives. Preferably, we would have assessed students' ability to apply Empirical methods to study Software Engineering through a small research project, similar to the design a study assignment.  
    However, the rules of Eindhoven Univrsity of Technology require  each course to contain an individual assessment. Therefore, grading cannot be based entirely on group projects. To include individual assessments, we could, of course, make the design of a study assignment individual. 
    However, because \Empirical is taught by the two authors of this chapter, it is unfeasible for them to give feedback on, and grade, individual assignments preserving an inherent diversity of individual research projects.
    Therefore, we combine the group project with an individual exam. }

\item \textcolor{black}{\emph{Ethics}: Ideally, we would allow students to use the full set of research methods discussed in the lectures and this includes research involving humans such as interviews or surveys. 
However, research with humans requires ethical approval from the Eindhoven ethics review board. Because of the course's limited time frame and the ethics review board's processing time, it is impossible to require students to obtain permission before executing their study. }

\end{itemize}

\section{Grades and Student Feedback}
In this section, we focus on the student assessment results, i.e., grades for Design a Study, Describe a Study, the exam, and the final course grade, as well as on the course evaluation by students.
While the former is intended to assess the quality of learning, the latter is intended to assess the quality of teaching, providing complementary perspectives on \Empirical. 
However, we know the limitations~\cite{Uttl2017Meta} and biases~\cite{Heffernan2022Sexism,Kreitzer2022Evaluating} inherent in student evaluations of teaching.
This is why we complement the formal student evaluation organized by the university with the feedback we have obtained from students during the course run through informal chats, dedicated feedback moments, anonymous feedback forms, and feedback obtained from colleagues.
We focus on the 2022 edition of the course and its exam that took place in January 2023, since the exam for the 2023 edition is scheduled for January 2024 and has not yet taken place at the time of writing.

\subsection{Grades}

\subsubsection{Describe a study and Design a study}
49 students have submitted the ``Describe a study'' assignment: four students have received less than 5.0, one student has received the perfect score of 10 and the median score was 7.1.

While working on the ``Design a study'' assignment, the students have been encouraged to work in groups of four. 
However, we did not want to enforce collaborations, and, hence some groups were smaller.
In the beginning of the course we had 9 four-student groups, 3 three-student groups, 3 two-student groups and 1 group consisting of a single student. 
Students from the smaller groups have been warned that the effort required to complete the assignment has been designed for four students and that the grading criteria will be applied to the submissions independently from the number of group members.
At the end 45 students have submitted the assignment: all students have passed the threshold of 5.0, and the median and the mean scores are 6.8. 

``Describe a study'' and ``Design a study'' have been designed to cover different learning objectives: \textbf{LO4} for ``Describe a study'' and \textbf{LO1}, \textbf{LO2}, \textbf{LO3}, and \textbf{LO5}.
This is why we expect the grades of the ``Describe a study'' and ``Design a study'' not to exhibit strong correlation.
This is indeed the case: Pearson's $r\simeq 0.015$, Spearman's $\rho \simeq = -0.1$, and none of these is statistically significant, suggesting that the two assignments indeed assess complementary aspects of the students' knowledge.

\subsubsection{Exam}
46 students have participated in the first exam attempt of the 2022 edition of \Empirical. 
Three students have submitted empty or near-empty exam papers and were clearly unprepared for the exam.
Six students have submitted a non-empty exam paper that could be seen as an attempt to pass the exam but got a grade lower than 5.0.
Seven students obtained a grade between 5.0 and 5.5: normally speaking, this grade would be seen as insufficient, but as explained in Section~\ref{sec:assessment},  with a high enough assignment grade, these students can still pass the course.  
Finally, the remaining 30 students have passed the exam.

\begin{figure}[t]
    \centering
    \includegraphics[width=0.9\linewidth]{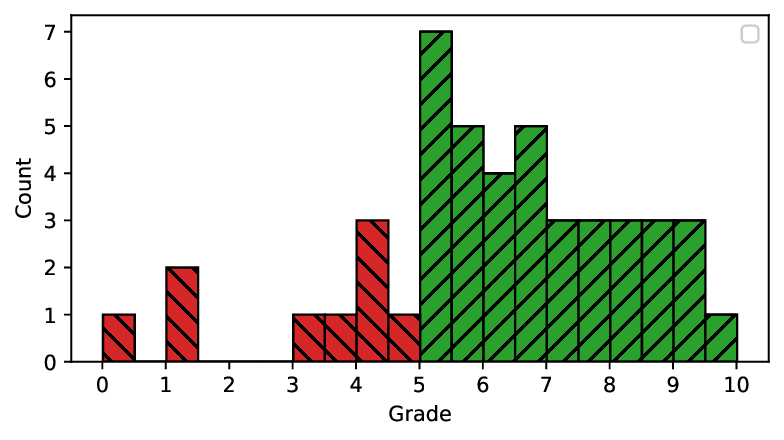}
    \caption{Histogram of the grade distribution for the exam.}
    \label{fig:enter-label}
\end{figure}

To understand the quality of the individual exam questions, we have used the analysis metrics included in ANS Delft, the assessment platform provided by our university.\footnote{\url{https://ans.app/landing}}
ANS Delft includes several such metrics; we opt for P and rir scores. 
The p-score is a measure of the difficulty of the question.
It is defined as the fraction of the participants who answered the question successfully. 
Meanwhile, the rir score computes the correlation between the grade of the question and the exam as a whole, with higher values indicating that students who performed well on the question, tend to receive a better grade.
The assumption underlying rir is that overall students performing well on the exam as the whole are expected to perform well on individual questions as well.

  \begin{figure}
    \centering
    \begin{subfigure}{0.49\linewidth}
      \includegraphics[width=0.65\linewidth]{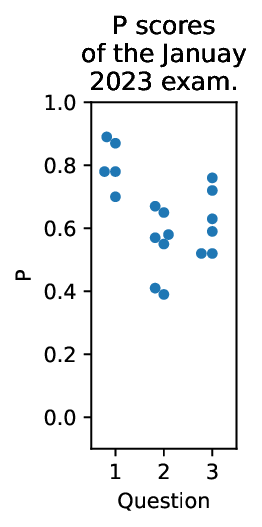}
      \caption{P-values as computed by ANS Delft}
      \label{fig:p-values}
    \end{subfigure}
\hfill
    \centering
    \begin{subfigure}{0.49\linewidth}
      \includegraphics[width=0.65\linewidth]{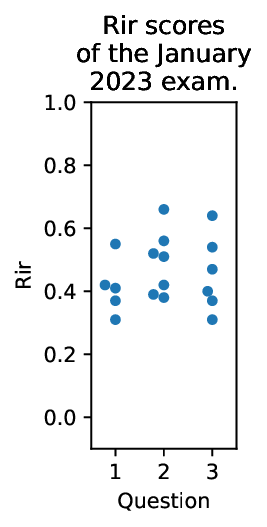}
      \caption{Rir values as computed by ANS Delft}
      \label{fig:rir-values}
    \end{subfigure}
    \caption{Quality of individual exam questions.} 
    \label{fig:exam:questions}
  \end{figure}

Figure~\ref{fig:exam:questions} summarizes the P-scores and rir score of the exam.
Each dot corresponds to the score of a subquestion and question to part of the exam.
For example, the five dots corresponding to Question 1 in Figure~\ref{fig:p-values} show P scores of five multiple-choice questions designed to check the understanding of the basic notions of empirical software engineering (see Section~\ref{sec:exam}).

Figure \ref{fig:p-values} shows that the P-scores are almost always higher than .4. There is only one P-score lower than .4, and its value is .39. 
The scores are higher for the subquestions of the first question than for the subquestions of the second and third questions. 
This is to be expected, as the closed subquestions for the first question are meant to be easier than the open subquestions for the second and third questions.
Meanwhile, the second and third questions are meant to be harder to answer, so the lower p-scores for these subquestions are expected and not immediately problematic. 
However, there are two subquestions for the second question that only 40\% of the students could answer correctly, which might be a sign to review these questions for future exams.

The Rir scores are plotted in Figure \ref{fig:rir-values}. 
There are no outliers, and most importantly, the rir values are distributed equally across the different questions. 
  The median value is slightly higher than .4, indicating that the outcome of most questions correlates with the outcome of the exam as a whole. 
  Overall rir scores higher than 0.4 are considered to be quite good.\footnote{\url{https://support.ans.app/hc/en-us/articles/360027234814-How-to-analyse-questions}} 
  
The high values for both metrics considered indicate that the exam is \textbf{reliable}.

\subsubsection{Course grades}
Out of 59 students registered for the course, 36 passed it, i.e., scored 6 or higher.
Unlike the assignment grades or exam grades, course grades are integers. 
8 students scored 6, 17 students scored 7, 9 students scored 8, and 2 students scored 9.
Among the remaining students, 16 did not complete the course requirements (i.e., did not submit homework assignments or did not participate in the exam).

\subsection{Feedback}

\subsubsection{Informal feedback}
As stated above, we complement the formal student evaluation organized by the university (see Subsection~\ref{sec:formalevaluation}) with the feedback we have obtained from students during the course run by means of informal chats, dedicated feedback moments, and anonymous feedback form, as well as feedback obtained from colleagues.

The set of concerns raised by students in the anonymous feedback form is very diverse. 
Students use it to ask questions. like whether we can upload slides to lectures before the lecture itself. 
However, they also use the feedback form to express their concerns: In both 2022 / 2023 and 2023 / 2024 students wrote that they were unsure about how to prepare for the exam. 
Students indicated that they found the lectures and primary material very abstract and theoretical and unsure how to answer the exam questions. 
We believe that the gap between theory and application of the theory is one of the challenges of teaching this empirical methods course: 
The currently available material is theoretical, and in a ten-weeks course it is challenging to give students enough opportunities to apply the material 

\subsubsection{Formal student evaluation}
\label{sec:formalevaluation}
Thirteen students have filled in the student evaluation form sent by the faculty. 
While this corresponds to a response rate of 22\% this response ratio is similar to response rates for other courses.
Overall, the students were positive: the course scored 3.8 on a scale of 5, and the teachers 4.2 on a scale of 5.
In particular, the students have agreed that the educational setup (e.g., structure, content, teaching/learning methods, level, and coherence) worked well and was suitable for this course; the course was well organized, and the course material was clear and motivated students to study.
The students expected that the material of this course would be useful for their graduation project, enjoyed the industrial talks at the start of the course, and believed the workshops were a useful addition to the course.
They further indicated that the lecturer explained the content clearly and comprehensively, while the instructor helped them master the subjects.

For us as teachers, the most valuable feedback was provided in the open questions---what they liked about the course and what could be improved.
The students mostly appreciated the relevance of the course topics (``Very applicable for anyone's master project and therefore important knowledge to have'') and their uniqueness in the educational offering of our computer science program, quality of the teaching materials (``The beautiful, clear slides'', ``notes with the slides were useful'', ``I liked how the slides were organized, they were creative and clear in structure''), diversity of the assignments (`it's an overview of many things I have studied/worked with before, so I could review everything again'', ``I enjoyed the topics that were discussed. They helped me better understand and classify studies I have been reading for other subjects or for my master's thesis.''), intermediate evaluation (``Iterative approach for submissions with opportunities for feedback along the way'', ``a lot of voluntary effort was put into intermediate feedback (ungraded assignments and coffee hours)''), and in general the teachers and the course atmosphere (``I liked specially the professors.'', ``great ability of instructors to generate real and impactful discussions, which helped me think critically about the topic.'', ``atmosphere in class was very nice, both lecturers were enthusiastic not only about the material but also about teaching it (not always a given in uni) and interacting with the class. $<$...$>$ the passion behind the course was definitely a big pro, i enjoyed attending the lectures'').
Students indicated that the non-graded character of the intermediate assignments did not motivate them and suggested determining at least 10\% of the grade based on the intermediate assignments.
We, however, believe that making these assignments graded would prevent the students from exploring (and making mistakes), which is crucial at early stages of learning.
Another comment referred to the feedback from Describe a study assignment and students' inability to apply the feedback for the non-graded submission to the graded one.
To address this comment in 2023 we have revised the Describe a study assignment as described in Section~\ref{sec:describe}.
Furthermore, one of the students requested a video recording of the lectures: while the video recordings of one of the pandemic course editions are available on YouTube, and the link to the recordings has been announced during the first lecture, we recognize that the quality of the videos is, of course, not professional. 
This is why we have requested a professional recording in 2023.
Finally, the students have indicated having difficulties with formulating research questions.
We recognize this challenge; however, this is an inherent challenge of conducting research, and learning how to formulate research questions is an important milestone in researchers' growth.
This is why we believe that this is an important goal of our course. 
To support the students' learning we provide detailed feedback on their homework, as well as discuss their research questions during one of the coffee hours. 

Based on the student evaluation the education program management has awarded us the Excellent Course Evaluation certificate.

\section{Lessons Learned and Didactic Challenges}

The lessons we have learned are partly related to the body of empirical software engineering knowledge, and partially to the specifics of our educational setting.

\subsection{Empirical Software Engineering Body of Knowledge}
While preparing for this course, we have observed that multiple articles are available on the application of individual research methods in the context of software engineering, e.g.,  interviews, surveys, or controlled experiments.
However, few articles provide a coherent vision on empirical software engineering and few articles are trying to focus on the steps preceding the choice of an individual research method, namely, formulating research questions and understanding what research questions can be answered by means of an empirical study.

The chapter by Easterbroook et al.~\cite{Easterbrook2008Selecting} makes an important step in supporting the learners in understanding how research questions can be formulated; however, having a well-formulated research question is not enough. 
A research question should also be \emph{relevant} for software engineering practice, research, or education, and it should be \emph{novel} with respect to the existing body of knowledge. 
Additionally, when formulating research questions we observed that some students focus very strongly on the data available in the datasets and formulate research questions that can be answered based on the datasets but lose sight of relevance or novelty. 
For example, focusing on the dataset of communication in Discord communities of  Python, Go, Racket, and Clojure~\cite{Subash2022Dataset} students suggest studying ``how often are programming languages different than the one corresponding to the channel, mentioned in a channel''. 
However, the report submitted did not indicate what stakeholder would benefit from answering this question and in what way(s).
Based on the feedback, the students decided to revise their question.
To support students in reflecting on the relevance and novelty of their research we have introduced them to the design science framework of Engström et al.~\cite{Engstrom2020How}, but this remains a challenge we will need to continue working in the years to come.
 
Another challenge we have observed is the absence of a coherent delineation of empirical research.
While all the sources seem to agree that empirical research should be based on observations, the further conceptualizations of this idea seem to diverge.
For example, the ABC framework~\cite{Stol2018ABC} and the comparison of rationalism and empiricism by Ralph~\cite{Ralph2018Two} focus on studies seeking knowledge, implicitly equating these notions.
Later on Stol and Fitzgerald~\cite{Stol2020Guidelines} (following Wieringa~\cite{Wieringa2009Design}) juxtapose knowledge-seeking and solution-seeking research.
While both knowledge-seeking and solution-seeking research can be empirical, the ABC framework would be appropriate for the former, framework of design science as discussed by Runeson et al.~\cite{Runeson2020Design} and Engström et al.~\cite{Engstrom2020How} would be appropriate for the latter.
However, the design science framework also includes studies focusing solely on problem understanding, even though these studies seek knowledge rather than solutions. 
Moreover, the ABC framework of empirical research also covers \emph{non-empirical} studies such as formal theory and computer simulation. 
This plurality of opinions and conceptualizations reflects the active quest for understanding inherent to scientific research, but at the same time, it makes it extra challenging to present the students with a coherent vision of empirical software engineering research. 

\subsection{Didactic challenges}
The next group of challenges are related to students being inexperienced researchers.

Conducting an empirical research study is by no means trivial and teaching research skills requires a lot of scaffolding.
We have provided significant support by offering intermediate ungraded assignments, providing datasets, and allocating dedicated sessions for students to ask questions beyond the regular teacher-student communication. 
\textcolor{black}{We would recommend working with students and giving them feedback throughout the course for anyone who wants to teach empirical methods to students. }

Next, we observed that students had a hard time using or acquiring technical skills related to analyzing data.
For example, while SQL is being taught in all Computer Science and Data Science bachelor programs around the world, and students are expected to have learned at some point during their studies, using this knowledge to answer questions about Stack Overflow using the Stack Exchange data explorer\footnote{\url{https://data.stackexchange.com/}} proved to be difficult.
Working with no-SQL databases such as MongoDB proved even more difficult, and we had to exclude datasets stored in such databases.
\textcolor{black}{Because the course focuses on empirical methods and not data analysis skills, we address this in the course design itself: We do not require any technical affinity, and we make available datasets to students that are easy to access.} 

Yet another challenge was related to the speed of reading academic texts. 
Since, unfortunately, reading academic papers is not part of a traditional bachelor curriculum in Computer Science, for many students \Empirical was the first exposure to academic writing.
This might be demotivating, as one of the students has indicated ``I am not used to research and reading papers as you are, no matter how much I practice it''.
Moreover, English is not the first language of our students.
As shown by Busby and Dahl, while on its own not being a native speaker does not necessarily reduce the reading speed of an academic text, combination of not being a native speaker and operating in a parallel language context where both English and the local language are used, might reduce the reading speed~\cite{Busby2021Reading}.
To address this challenge, we have explained to the students how empirical papers are structured and encouraged them to highlight the relevant parts of a paper when reading it.
We have also consciously limited the mandatory reading to four-page papers.

Finally, we have observed that many students were afraid of working with qualitative analysis approaches.
This complex phenomenon can partly be attributed to the strong positivist tradition in computer science in general and in fundamental courses taught in the Bachelor of Computer Science program in particular.
We are, of course, not the first ones to observe this---Richards has discussed her experiences with teaching qualitative research course to students with the positivist backgrounds and our experiences to some extent echo hers~\cite{Richards2011Every}.
Two important differences, however, make our settings different from the one discussed by Richards, and simultaneously limit our ability to address the ``qualitative analysis fear''.
First, qualitative analysis is merely a subject of slightly more than one lecture in a single course in the master degree.
As such it remains ``the odd one out'', and students do not get enough time and opportunity to learn more about qualitative research and better appreciate it.
Second, while both of the authors have conducted qualitative research as part of their empirical software engineering work, both of us are trained as Computer Scientists and had to learn qualitative analysis by self-learning and in an engineering context rather than through an established academic curriculum in a scientific discipline with a long tradition of qualitative inquiry.
\textcolor{black}{We unfortunately have no practical recommendation on how to address this lesson learned. Given the limitations of the course and especially the time frame, there is no time to give students more hands-on experiences with qualitative research.}

\section{Resources}
Both the ``pandemic'' video recordings of the 2020 edition\footnote{\url{https://www.youtube.com/@ads4se464}} and the classroom video recordings of the 2023 edition\footnote{\url{https://videocollege.tue.nl/Mediasite/Channel/c92d73ba0bf940ae830945ed18ec0ba95f}} are publicly available. We also make slides available as .pdfs, previous exams including answer keys, and the syllabus. These can be accessed through the repository with supplementary materials accompanying this book.\footnote{\url{https://zenodo.org/doi/10.5281/zenodo.11544897}}

\section{Conclusions}

In this chapter, we have described the \Empirical course taught by us at Eindhoven University of Technology. We have aimed to provide a rich description of the challenges we have experienced and decisions taken to address them. We further elaborate on the structure of the course and assessment strategy we implemented and reflect upon the feedback received from students and peers. To support colleagues teaching similar courses elsewhere we share a plethora of teaching materials, from lecture recordings and slide decks to homework assignments and assessment rubrics. We are mindful that teaching at a different institution might require radically different approaches but hope that our materials and insights will be able to serve as stepping stones.

\bibliographystyle{plain} 
\bibliography{references} %
\end{document}